# Digital-assisted photonic analog wideband multipath self-interference cancellation

Moxuan Han[a,b], Taixia Shi[a,b], and Yang Chen[a,b,*]

[a] Shanghai Key Laboratory of Multidimensional Information Processing, East China Normal University, Shanghai, 200241, China
[b] Engineering Center of SHMEC for Space Information and GNSS, East China Normal University, Shanghai, 200241, China
[*] ychen@ce.ecnu.edu.cn

**ABSTRACT**
A digital-assisted photonic analog wideband radio-frequency multipath self-interference cancellation (SIC) and frequency downconversion method based on a dual-drive Mach–Zehnder modulator and the recursive least square (RLS) algorithm is proposed and demonstrated for in-band full-duplex systems. Besides the reference for the direct-path self-interference (SI) signal, the RLS algorithm is used to construct another reference for the residual SI signal from the direct path and the SI signals from the reflection paths. The proposed method can solve the performance limitation in the previously reported SIC methods of constructing the multipath SI signal using a single reference caused by the limited dynamic range of the digital-to-analog converter when the direct-path SI signal is much stronger than the sub-weak reflection-path SI signals. An experiment is performed. When the carrier frequency of the multipath SI signal is 10 GHz and the direct-path SI signal is much stronger than the sub-weak multipath SI signal, the cancellation depths of about 26.7 and 26.1 dB are realized with SI baud rates of 0.5 and 1 Gbaud. When the direct-path SI signal and sub-weak multipath SI signal own closer power, the corresponding cancellation depths are 24.7 and 20.8 dB, respectively.

## 1. Introduction

With the extensive use of radio-frequency (RF) communication devices, wireless spectrum resources have become scarce natural resources. Therefore, how to improve spectrum utilization has become an urgent problem to be solved. One of the solutions is using the in-band full-duplex (IBFD) communication technique, which can double the spectrum efficiency in theory [1]–[3]. Different from the conventional time-division duplex and frequency-division duplex systems, the IBFD system allows signals to be transmitted and received in the same frequency band simultaneously. However, due to the co-allocation of the transmitting antenna (TA) and the receiving antenna (RA), the signal transmitted by the TA can also be received by the RA and becomes the self-interference (SI) signal. The SI signal is much stronger than the signal of interest (SOI) received by the RA so that the SOI is submerged in the SI signal in both spectral and temporal domains. Because the SI signal and the SOI are in the same frequency band, it is impossible to separate them using an electrical bandpass filter.

To solve this problem in the IBFD system, many electrical-based self-interference cancellation (SIC) methods have been studied [4], mainly including antenna domain cancellation [5], analog domain cancellation [6], and digital domain cancellation [7]. These three methods are usually combined to achieve a good cancellation result. However, due to the electronic bottleneck, the electrical-based SIC methods are limited in operating frequency and bandwidth. Photonic-assisted SIC methods [8]–[10] have been widely studied in recent years because of the inherent advantages of microwave photonics, such as low loss, wide bandwidth, and high operating frequency. However, most photonic and optoelectronic hybrid SIC systems only focus on the cancellation of the line-of-sight SI signal without considering the influence of the multipath effect. When the multipath SI signal is taken into account, the system usually requires multiple analog links and a large number of delay and attenuation devices or a large number of laser diodes [11]–[13]. To reduce the number of devices, simplify the SIC system, and improve the effectiveness of the SIC, the digital adaptive filter is introduced to reconstruct a single reference signal for multipath SI signal in the analog domain to cancel the SI signal using a single analog link. In [14], the least mean square algorithm is introduced to the optical self-interference cancellation (OSIC) system to reconstruct the reference signal, while in [15], the least square algorithm is used to implement an adaptive frequency-domain pre-equalizer to suppress the multipath SI signal. However, in [14] and [15], the influence of the limited dynamic range (DR) of the digital-to-analog converter (DAC) on the cancellation performance of multipath SI signal is not considered. Multipath SI signal usually consists of a strong direct-path SI signal, some sub-weak reflection-path SI signals, and some ultra-weak reflection-path SI signals (comparable to the SOI). In [14], [15], the multipath channel is directly estimated, and the reference signal for the multipath SI signal is directly generated. Due to the DAC quantization noise, when the direct-path SI signal is much stronger than the sub-weak reflection-path SI signals in power, the limited DR of the DAC becomes a performance-limiting factor, especially when the number of DAC bits is small [16].

In this letter, a digital-assisted photonic analog wideband multipath SIC and frequency downconversion method is proposed and experimentally verified to eliminate the strong direct-path SI signal and the sub-weak reflection-path SI signals. Compared with the methods in [14], [15], an analog reference only for the direct-path SI signal is employed to cancel the strong direct-path SI signal, while another analog reference for the residual direct-path SI signal and the sub-weak reflection-path SI signals is employed to realize multipath analog SIC, therefore solving the problem that the limited DR of the DAC leads to poor SIC performance when the direct-path SI signal is much stronger than the sub-weak reflection-path SI signals. An experiment is performed. When the carrier frequency of the multipath SI signal is 10 GHz and the direct-path SI is much stronger than the multipath SI signal, with the digital-assisted analog cancellation, the cancellation depths of about 26.7 and 26.1 dB are realized with baud rates of 0.5 and 1 Gbaud. When the direct-path SI and multipath SI signal own closer power, the corresponding cancellation depths are 24.7 and 20.8 dB, respectively.

**2. Principle**

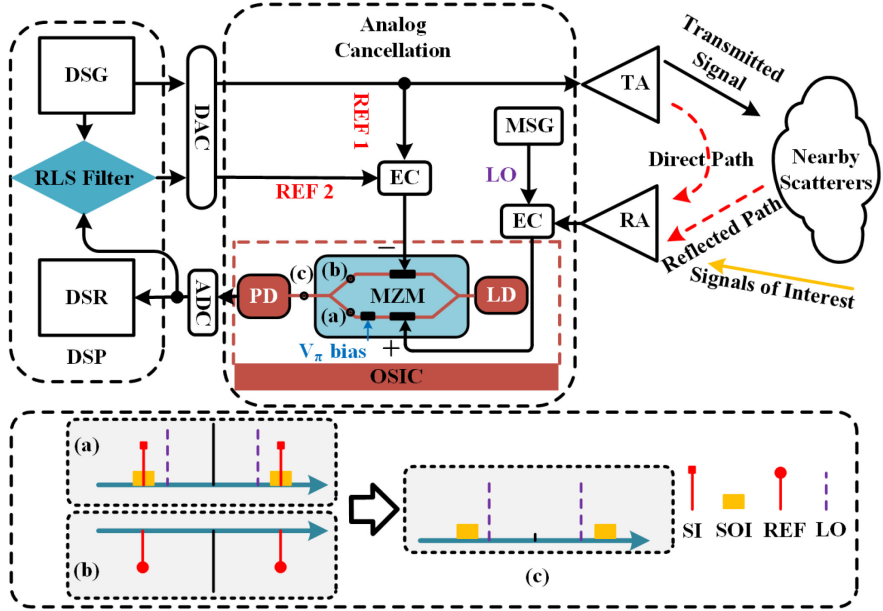

Fig. 1. The schematic diagram of the proposed digital-assisted analog SIC and frequency downconversion system. DSG, digital signal generation; DSR, digital signal reception; DSP, digital signal processing; ADC, analog-to-digital converter; DAC, digital-to-analog converter; EC, electrical coupler; TA, transmitting antenna; RA, receiving antenna; REF, reference; OSIC, optical self-interference cancellation; MSG, microwave signal generator; LO, local oscillator; MZM, Mach–Zehnder modulator; LD, laser diode; PD, photodetector. (a)-(c) are the diagrams of the spectra at different locations in the system.

The schematic diagram of the proposed system is shown in Fig. 1. The transmitted signal from the digital signal processing (DSP) unit is converted into an analog signal by a DAC and then divided into two parts: one part is sent to the TA as the transmitted signal and the other part is used as a reference signal (REF1) for the direct-path SIC. The transmitted signal is received by the RA via the direct and reflection paths along with the SOI. After the OSIC system, the direct-path SI signal is much suppressed by REF1, and the residual direct-path SI signal and the sub-weak reflection-path SI signals need to be further canceled. The received signal after OSIC is converted back to the electrical domain and sampled by an analog-to-digital converter (ADC). The digitized signal is processed by the recursive least square (RLS) algorithm to construct the reference signal (REF2) for the residual direct-path SI signal and the sub-weak reflection-path SI signals. REF2 is finally converted to the analog domain via the DAC and combined with REF1 for the multipath SIC.

The OSIC method used in this letter is an evolution of the method we have reported in [10]. In this work, a local oscillator (LO) signal is added to downconvert the signal to the intermediate frequency (IF) band along with the SIC operation to facilitate the digital domain signal processing.

For the RLS filter, it works by updating the parameter $\hat{W}(n)$ of the filter to minimize the cumulative mean square error, which can be expressed as

$$\varepsilon(n) = \min \sum_{k=0}^{n} \lambda^{n-k} e^2(k), \tag{1}$$

where $\lambda$ is the forgetting factor of the RLS filter, $e(k)=d(k)-\hat{W}^H(k-1)x(k)$, $d(k)$ is the expected signal of the RLS filter, and $y(k)=\hat{W}^H(k-1)x(k)$ is the output signal of the RLS filter. To minimize $\varepsilon(n)$, $\hat{W}(n)$ is updated according to

$$\hat{W}(n) = \hat{W}(n-1) + k(n)e^*(n), \quad (2)$$

where $k(n)$ can be expressed as

$$k(n) = \frac{\lambda^{-1}P(n-1)x(n)}{1+\lambda^{-1}x^H(n)P(n-1)x(n)}. \quad (3)$$

Here, $x(n)$ is the input signal of the RLS filter and $P(n)=\lambda^{-1}P(n-1)-\lambda^{-1}k(n)x^H(n)P(n-1)$. The expected signal $d(n)$ of the RLS filter in this system is the residual direct-path SI signal and the sub-weak reflection-path SI signals S(*residual SI*) digitized by the ADC, i.e., $d(n)=S(residual\ SI)$; the input signal $x(n)$ is the original SI signal S(*original*) generated in the digital signal generation (DSG), i.e., $x(n)=S(original)$; the output signal $y(n)$ is the reference signal S(*match*) of the residual direct-path SI signal and the sub-weak reflection-path SI signals, i.e., $y(n)=S(match)$. The RLS adaptive filter is divided into three parts as follows:

Step 1: Disable the SOI, the digital signal generated by the DSG is transmitted out by the TA and becomes a multipath SI signal at the RA after free-space transmission. Then, the multipath SI signal is canceled by REF1 at the OSIC.

Step 2: The S(*residual SI*) and S(*original*) are used as $d(n)$ and $x(n)$ of the RLS filter, respectively. The parameter $\hat{W}(n)$ of the filter can be obtained by iterating through the RLS algorithm, and the S(*match*) is constructed by the RLS filter.

Step 3: S(*match*) is analogized by the DAC to become REF2, and REF1 and REF2 are coupled to cancel out the residual direct-path SI signal and the sub-weak reflection-path SI signals in the OSIC system.

## 3. Experimental results and discussion

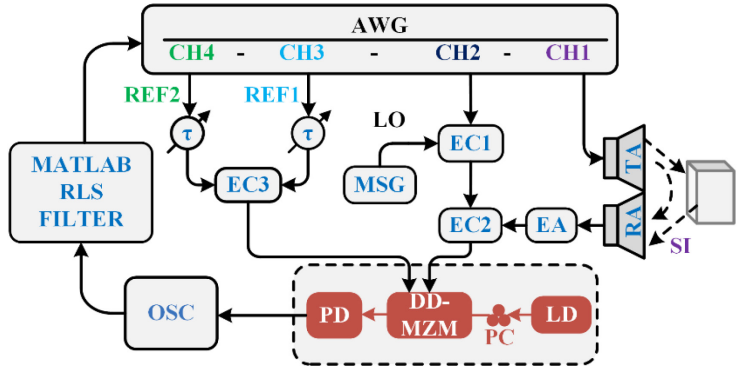

Fig. 2. Experimental setup of the proposed system. CH, channel; TA, transmitting antenna; RA, receiving antenna; EC, electrical coupler; MSG, microwave signal generator; EA, electric amplifier; LD, laser diode; PC, polarization controller; DD-MZM, dual-drive Mach–Zehnder modulator; PD, photodetector; τ, tunable electrical delay line; AWG, arbitrary waveform generator; OSC, oscilloscope; LO, local oscillator; SI, self-interference.

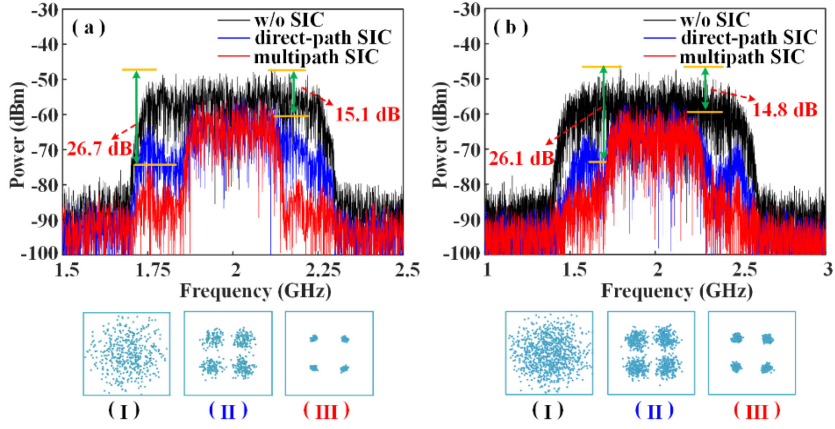

Fig. 3. Electrical spectra and constellation diagrams of the downconverted IF signal without SIC, with direct-path SIC, and with multipath SIC. The SI signals are from two paths, the carrier frequency of the signals is 10 GHz, and the baud rate of the QPSK-modulated SI signal is (a) 0.5 Gbaud, (b) 1 Gbaud. (I)-(III) SOI constellation diagrams.

The proposed system is experimentally demonstrated based on the setup shown in Fig. 2. The SI signal (also the transmitted signal), SOI, REF1, and REF2 are generated by an arbitrary waveform generator (AWG, Keysight M8195A, 64 GSa/s, 8bit). The transmitted signal is radiated to the free space using a TA (GHA080180-SMF-14, 8-18 GHz), whereas the SI signals from the direct path and the reflection paths are received using an RA (GHA080180-SMF-14, 8-18 GHz). The reflector is placed very close to the antennas, so the transmitted signal is direct-fed to the TA with no amplification. Note that, in the experiment, the SOI is added along with an LO signal via EC1 and EC2 after the SI signal is received by the RA and amplified by an electrical amplifier (EA, ALM 145-5023-293 5.85-14.5 GHz), instead of directly received by the RA. A 14.5-dBm CW lightwave generated from a laser diode (LD, ID Photonics CoBriteDX1-1-C-H01-FA) with a wavelength of 1550.93 nm is injected into a dual-drive Mach–Zehnder modulator (DD-MZM, Fujitsu FTM7937EZ200) via a polarization controller (PC). The DD-MZM is biased at the minimum transmission point and the lower arm of the DD-MZM is driven by the signal from EC2 consisting of the multipath SI signal, the SOI, and the LO signal. The 15-dBm LO signal with a frequency of 8 GHz is generated from a microwave signal generator (MSG, Agilent 83630B). The upper arm of the DD-MZM is driven by the signal from EC3 (Narda 4456-2) consisting of REF1 and REF2. The optical signal from the DD-MZM is detected in a photodetector (PD, Nortel PP-10G) and the photocurrent from the PD is converted into a digital signal through an oscilloscope (OSC, R&S RTO2032). At the beginning of the iteration, the parameter $W(0)$ and forgetting factor $\lambda$ are set to $W(0)=0$ and $\lambda=1$, while the filter order is set to 160 and the 160-by-160 matrix is set to $P(0)=I$. Then, the RLS algorithm is applied to obtain the RLS adaptive filter coefficients, and then REF2 for the residual direct-path SI signal and the reflection-path SI signal is constructed and generated from CH4 of the AWG. Although the amplitude and delay of the references are adjusted in the digital domain, a further delay fine-tuning step is achieved by using two electrical delay lines due to the limited tuning step of the AWG (15.6 ps).

Firstly, the system is verified when a strong direct-path SI signal and a sub-weak reflection-path SI signal having a power difference of 24 dB are received. The carrier frequency of the SOI and SI signals is 10 GHz, the frequency of the LO signal is 8 GHz, and the baud rate of the SI is 0.5 or 1 Gbaud. To show the SIC more intuitively, the baud rate of the SOI is set to half of the SI baud rate. The signals are all quadrature phase-shift keying (QPSK) signals. As shown in Fig. 3, after the direct-path SIC, the SI signal is suppressed by about 15.1 and 14.8 dB when the baud rates of the SI are 0.5 and 1 Gbaud, respectively. After the direct-path analog cancellation, the residual direct-path SI signal and the reflection-path SI signal interfere with each other and some ripples are observed in the spectrum. Then, the analog SIC is further implemented by constructing REF2 for the residual direct-path SI signal and the reflection-path SI signal via the RLS algorithm. As can be seen, after the digital-assisted analog cancellation, total cancellation depths of about 26.7 and 26.1 dB are realized for the SI signals with baud rates of 0.5 and 1 Gbaud. In addition, the constellation diagrams of the SOIs are also shown in Fig. 3. When only the direct-path SIC is implemented, the constellation changes from chaos to four constellation points, which are basically recognizable. The corresponding error vector magnitudes (EVMs) of the QPSK SOI signals are 33.3% and 35.2% in Fig. 3(a) and (b), respectively. When multipath analog SIC is implemented, the EVMs are further improved to 9.1% and 13.4%, respectively.

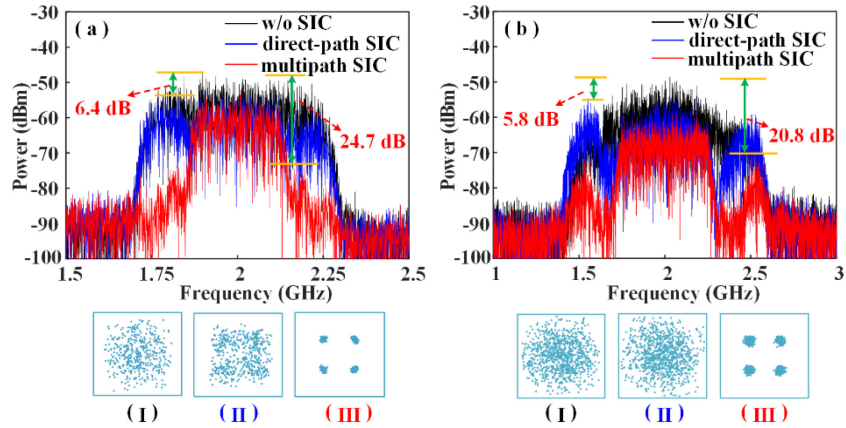

Fig. 4. Electrical spectra and constellation diagrams of the downconverted IF signal without SIC, with direct-path SIC, and with multipath SIC. The SI signals are from three paths, the carrier frequency of the RF signal is 10 GHz, and the baud rate of the QPSK-modulated SI signal is (a) 0.5 Gbaud, (b) 1 Gbaud. (I)-(III) SOI constellation diagrams.

Then, the SIC system is further verified with a direct-path SI signal and two sub-weak reflection-path SI signals. In this case, the power of the three signals is set to be closer (2 and 6 dB power differences) via placing the reflector properly in front of the antennas. The experimental results are shown in Fig. 4. Because the SI signals have closer power, the cancellation depths after the direct-path analog SIC are 6.4 and 5.8 dB for the SI signals with baud rates of 0.5 and 1 Gbaud, which is much smaller than those in Fig. 3. The constellation diagrams are still indistinguishable after the direct-path analog SIC. However, after the following multipath analog SIC, total cancellation depths of 24.7 and 20.8 dB are achieved, which is very close to those shown in Fig. 3. The constellation diagrams are also distinguishable after the multipath analog SIC. In this case, the EVMs of the SOIs are 9.4% and 13.9%, respectively.

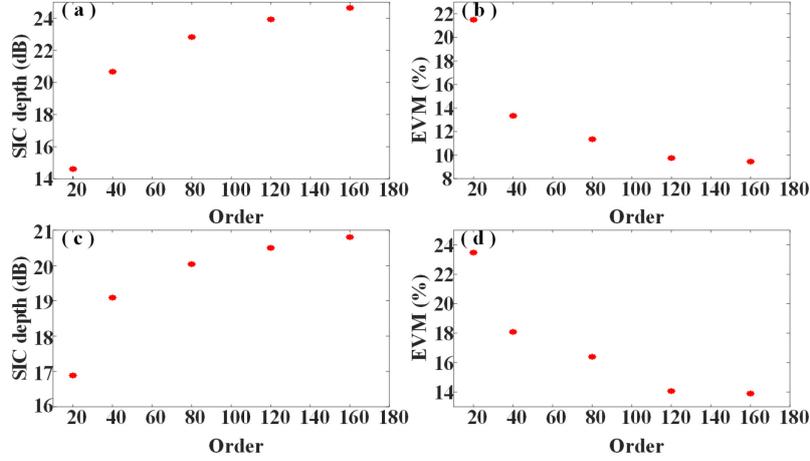

Fig. 5. The variation of the cancellation depth of the SI signals and the EVM of the IF signals using RLS filters with different orders. The baud rate of the SI signal is (a), (b) 0.5 Gbaud and (c), (d) 1 Gbaud.

Finally, the influence of the order of the RLS filter on the cancellation depth of the SI signals and the EVM of the SOIs is studied. As can be seen from Fig. 5, the cancellation depth increases with the increase of the order of the RLS filter, while EVM decreases on the contrary. It is also indicated from Fig. 5 that when the order of the RLS filter is small, the SIC effect can be enhanced with the increase of the order. However, when the filter order exceeds 120, the performance improvement introduced by the increase of the filter order decreases. This is because, in this case, the matching of the delay and amplitude and other distortions of the system instead of the order of the filter become the main reason to limit the cancellation depth of the SI signals. In addition, when the filter order increases, the amount of computation also increases, which inevitably slows down the iterative speed of the RLS algorithm. Therefore, it is important to select a proper filter order by comprehensively considering the convergence time and SIC depth.

Limited by the DR of the 8-bit ADC in the OSC, the SOI is not set much smaller than the SI signals to ensure that the SOI can be captured. Thus, in the experiment, the RLS filter works without SOI. In real-world IBFD systems, the SOI is indeed much smaller than the strong direct-path SI signal and the sub-weak reflection-path SI signals, and the influence of the SOI on the RLS filter can be ignored. In addition, the residual SI signals after the analog domain SIC and the ultra-weak SI signals that are comparable with the SOI in power can be further processed in the digital domain to achieve a deeper cancellation depth.

## 4. Conclusion

A digital-assisted photonic analog multipath SIC and frequency downconversion approach for the IBFD system is proposed and experimentally demonstrated. The key significance of the work is that the multipath analog SIC is implemented by using two reference signals including a reference only for the direct-path SI signal and a reference for the residual direct-path SI signal and the sub-weak reflection-path SI signals, therefore solving the problem that the limited DR of the DAC leads to poor SIC performance when the direct-path SI signal is much stronger than the reflection-path SI signals. The wideband multipath cancellation capability using a QPSK-modulated signal with the baud rates up to 0.5 and 1 Gbaud is demonstrated, and cancellation depths of more than 26.7 and 26.1 dB are achieved.

**Acknowledgements**

This work was supported by the National Natural Science Foundation of China [grant number 61971193]; the Natural Science Foundation of Shanghai [grant number 20ZR1416100]; the Open Fund of State Key Laboratory of Advanced Optical Communication Systems and Networks, Peking University, China [grant number 2020GZKF005].